\documentstyle[prl,aps,psfig,graphicx,multicol]{revtex}

\def\be{\begin{equation}}
\def\ee{\end{equation}}
\def\ba{\begin{eqnarray}}
\def\ea{\end{eqnarray}}

\begin{document}

\title{Paramagnetic Instability at Normal-Metal -- Superconductor 
Interfaces}

\author{Alban L.\ Fauch\`ere$^a$, Wolfgang Belzig$^b$, 
and Gianni Blatter$^{a}$}

\address{$^a$Theoretische Physik, Eidgen\"ossische Technische Hochschule,
 CH-8093 Z\"urich, Switzerland\\ 
$^b$Institut f{\"u}r Theoretische Festk{\"o}rperphysik, Universit{\"a}t
  Karlsruhe, D-76128 Karlsruhe, Germany}
\date{December 15, 1998}
\maketitle

\begin{abstract}
  We study the proximity coherence in a mesoscopic normal-metal film
  (N) in contact with a superconductor (S). Accounting for a repulsive
  interaction between the electrons in the normal metal, we find an
  enhanced local density of states close to the NS interface. The
  sharp peak in the density is pinned to the Fermi energy and
  leads to spontaneous paramagnetic interface currents.  The induced
  orbital magnetic moments exhibit the characteristic features of
  paramagnetic reentrance observed in normal-metal coated
  superconducting cylinders [Phys.\ Rev.\ Lett.\ {\bf 65}, 1514
  (1990)].
\end{abstract}

\begin{multicols}{2}
  
A normal metal in contact with a superconductor exhibits the
phenomenon of proximity --- the superconductor exports its coherent
state across the interface into the normal metal. On a microscopic
level, this phenomenon is described through the Andreev reflection
of the normal-metal quasi-particles at the
normal-metal--superconductor (NS) interface, converting normal- to
supercurrent. Proximity superconductivity exhibits a rich
phenomenology and has attracted considerable interest
recently\cite{curacao}.  A particularly puzzling finding is the
ultra-low-temperature reentrance observed in
normal-metal coated superconducting cylinders \cite{motaprl}, where,
contrary to expectation, the fully diamagnetic cylinder develops a
paramagnetic response at low temperatures.  Recently, it has been
speculated that a novel kind of persistent current states circling 
the cylinder might be responsible for this phenomenon\cite{bi}, 
but closer inspection
of the experimentally measurable quantities reveals that the
predicted effect is too small\cite{fgb}. In 
this Letter, we demonstrate that the presence of a repulsive
electron-electron interaction in the normal metal naturally leads to 
the appearance of a paramagnetic instability at very low temperature, 
offering a possible explanation of the reentrance effect in the NS 
cylinders. 

\begin{figure}[htbp]
\psfig{figure=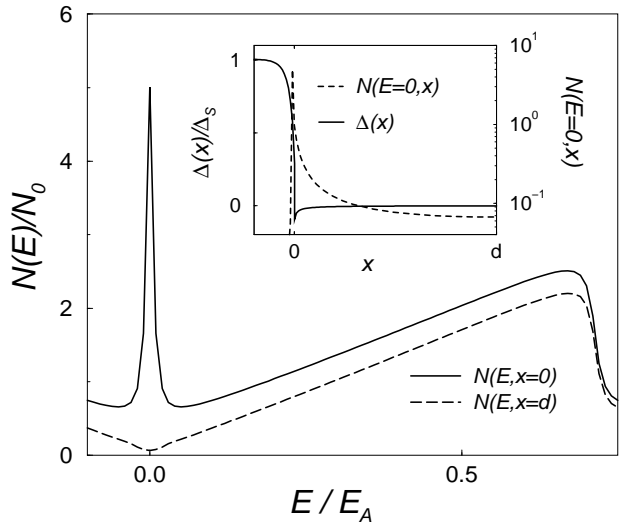,width=9.5cm,angle=-90}
\narrowtext
\vspace{0.3cm}
\caption{Local DOS $N(E,x)$ at the NS interface $x=0$ and 
at the metal boundary $x=d$ ($E_A=\hbar v_F/d$), 
as it follows from the self-consistent
solution of the Eilenberger equation, Eqs.\ 
(\ref{eilenberger}) and (\ref{gapeq}), for a thickness  
$d=10 \hbar v_F/\Delta_S$ and the coupling constants $V_S=-0.3$ and 
$V_N=0.1$. Inset: Spatial dependence of the order parameter $\Delta(x)$ 
and local DOS $N(E=0,x)$ at the peak energy (S: $x<0$, N: $x>0$).}
\label{dosfig}
\end{figure}

To be specific, we shall consider a clean normal-metal slab of
thickness $d$ ($0<x<d$), in perfect contact with a bulk, conventional
superconductor. The proximity effect is mediated by the Andreev
reflection at the interface with the superconductor, which transforms
incident electrons into back-reflected holes, thus binding the
quasi-particles states to the normal layer for $E<\Delta_S$.  In the
usual free electron gas description of the normal metal, the Andreev
bound states are found at $E_n = \hbar v_x (2n+1)\pi /4d$ ($n=0, 1,
..$; $v_x=v_F\cos\vartheta$) producing a linear suppression of the
DOS\cite{james} $N(E)\sim N_0 Ed/\hbar v_F$ close to the Fermi level
$E=0$ ($N_0=mk_F/\hbar^2\pi^2$). In the following we assume that the
electron-electron interaction in the normal layer, which follows from
the delicate balance between the phonon-mediated- and the
Coulomb-interaction, is repulsive. As a consequence, a finite order
parameter $\Delta(x)$ is induced in the metal, opposite in sign as
compared to $\Delta_S$ in the superconductor, see Ref.\ 
\cite{deGennes}.  The NS junctions then behaves like a Josephson
junction with a phase difference $\pi$, trapping quasi-particle states
at the Fermi energy close to the NS interface. The local density of
states $N(E,x)$ exhibits a peak at zero energy on top of the Andreev
density of states, as shown in Fig.\ \ref{dosfig}.  This peak involves
a macroscopic number of states with density $n_p\sim k_F^2/d$, which
in the following we call the $\pi$-states.

The change in the DOS crucially affects the response of the proximity
metal.  The linear current response $j[A]$ can be divided into two
contributions $j=j_{\rm dia}+j_{\rm para}$, the diamagnetic current
$j_{\rm dia}=-(e^2n/mc)\, A$ giving the rigid response of the bulk
density $n=k_F^3/3\pi^2$ and the paramagnetic current $j_{\rm para}$
following from the deformation of the wavefunction at the Fermi
surface\cite{schrieffer}, 
\be j_{\rm para}= \frac{e^2n}{mc}\, A \int
dE \left(-{ \frac{\partial f} {\partial E}}\right) \frac{N(E)}{N_0} 
\label{paraestimate} 
\ee 
for slowly varying fields $A$ ($f$ is the Fermi occupation
number).  While in a bulk superconductor the paramagnetic current is
quenched by the energy gap at low temperatures producing a net
diamagnetic response, the paramagnetic current of a bulk normal metal
cancels the diamagnetic current exactly. In the non-interacting metal
under proximity, the linear density of states suppression $N(E)\propto
E$ is still sufficient to suppress the paramagnetic current at zero
temperature\cite{zaikin}. Including a repulsive interaction places the
system in the opposite limit: The sharp DOS peak at the Fermi level
produces a paramagnetic signal which {\it over}compensates the
diamagnetic response. Such a paramagnetic response naturally leads to
an instability: The free energy $\delta F = -c j \delta A <0$ can be
lowered via a non-zero magnetic induction induced by spontaneous
currents along the NS interface. The interface currents are associated
with an orbital magnetization $M(T)$ producing a low-temperature
reentrance in the magnetic susceptibility.

In the following we present a quantitative analysis of the
paramagnetic instability induced by the $\pi$-states. The magnetic
induction $B_z(x)$ parallel to the surface is described by the vector
potential $A_y(x)$ which drives the currents $j_y(x)$. The
electron-electron interaction in the superconductor is accounted for
by an effective coupling constant $V_S<0$ and similarly $V_N>0$ in the
normal metal, see also Refs.\ \cite{nazarov,spivak}. Two
self-consistency problems have to be solved: First, we evaluate the
order parameter $\Delta(x)$ accounting for the different coupling
constants in the superconductor and the normal metal, and obtain the
local DOS $N(E,x)$.  Second, we determine the current functional
$j[A]$ which we solve together
with Maxwell's equation to find the spontaneous interface currents.\\
\vspace{-1.7cm}

We use the quasi-classical description following from the Eilenberger
equation for $\hat{g}$, 
\be -\hbar v_x \partial_x \hat{g} =\left[
  \{\hbar \omega +ie v_y A_y(x)/c\} \hat{\tau}_3 + \Delta(x)
  \hat{\tau}_1, \hat{g} \right],
\label{eilenberger}
\ee
where the 2$\times$2 matrix $\hat{g}$ contains the Green's functions 
$g_{\omega}(x, v_x)$ and $f_{\omega}(x,v_x)$ 
($\omega$ Matsubara frequency, $v_x=v_F\cos\vartheta$, $\hat{\tau}_i$ Pauli 
matrices, see Ref.\ \cite{bbf}). Eq.\ (\ref{eilenberger}) is completed
by the self-consistency relation for the pair potential 
($\langle .. \rangle$ is the angular average),
\be \Delta(x) = - V N_0 \pi T \sum_{\omega>0} \langle f_{\omega}(x,v_x) 
\rangle.
\label{gapeq}
\ee
The self-consistent numerical solution
of Eqs.\  (\ref{eilenberger}) and (\ref{gapeq}) is shown in the inset of
Fig.\ \ref{dosfig}. 
The course of the order parameter in the normal layer is asymptotically 
given by $\Delta(x) \sim -V_N N_0 \hbar v_F/x$, as expected from the 
$f$-function in the non-interacting case $V_N=0$. 
$\Delta(x)$ decays from a value $\sim -|V_N/V_S| \Delta_S$ at 
the NS interface, to $\sim  -V_N N_0 \hbar v_F/d$ at the outer
boundary. Close to the NS interface, the local DOS $N(E,x)=N_0 
{\rm Re} [\langle g_{-iE+\delta}(x,v_x)\rangle ]$ 
exhibits a pronounced peak at zero energy, as shown in Fig.\ \ref{dosfig}.
In order to proceed with analytical results, we approximate the order 
parameter by a step function,
\[ \Delta(x) = \left\{ \begin{array}{ll} \Delta_S, & x<0, \\ 
-\Delta_N, \quad\quad& 0<x<d, 
\end{array} \right. 
\]
where $\Delta_N \propto V_N$ enters as a parameter. The Green's function 
in the normal layer $x>0$ can be determined exactly and takes the form
\be g_\omega(x,v_x) = 
\frac{\hbar \omega \sinh \left[ \chi(d)-\gamma\right] + 
\Delta_N \cosh \left[ \chi(d-x)\right] }
{\hbar \Omega \cosh \left[\chi(d)-\gamma\right]},
\label{green_finite_d}
\ee
where $\chi(x) = 2 \Omega x / v_x$, $\hbar^2 \Omega^2 = \Delta_N^2 + 
\hbar^2\omega^2$, and $\tanh \gamma=\Delta_N/\hbar \Omega$ 
(we consider the limit $\Delta_S \gg \Delta_N, \hbar\omega$).
The second term in (\ref{green_finite_d}) describes the $\pi$-states 
at the NS interface. 
The poles of the Green's function at $\hbar\omega=-iE+0$ yield the 
bound state energies. While for $E\gg \Delta_N$ 
the Andreev states of the free electron gas are down-shifted by 
$\delta E_n \approx - 2\Delta_N/(2n+1)\pi$ ($n=0, 1,.. $),
below the gap $E<\Delta_N$ we find the $\pi$-states at 
\be E = \Delta_N / \cosh \frac{2\sqrt{\Delta_N^2-E^2}d}{\hbar v_x} 
\sim \Delta_N {\rm e}^{-2\Delta_N d/ \hbar v_x},  
\label{energy_zbs} 
\ee
exponentially close to Fermi energy. All trajectories with 
$v_x=v_F\cos\vartheta \ll \Delta_N d/\hbar$ possess a bound state 
at $E\approx 0$, thus producing the macroscopic weight of the zero 
energy DOS peak: For $\Delta_N > \hbar v_F/d$ the number of $\pi$-states 
per unit surface $N_{\rm surf}$ is equal to the number of transverse 
levels $N_{\rm surf} \sim k_F^2$, while for $\Delta_N < \hbar v_F/d$ 
it is reduced to $N_{\rm surf} \sim k_F^2 (\Delta_N d/\hbar v_F)^2$ 
via the reduction of the available solid angle $\cos\vartheta < 
\Delta_N d/\hbar v_F$.

We derive the current-field relations at low temperatures, assuming 
$k_B T \ll \hbar v_F/d, \Delta_N$. This implies a thermal length 
$\xi_N(T) = \hbar v_F/2\pi k_B T$ larger than the thickness $d$ and no 
thermal smearing on the scale $\Delta_N$. Only the trajectories with
$\cos\vartheta < \Delta_N d/\hbar v_F$ contribute to the current 
at low temperatures. We describe them in the limit 
$\hbar v_x/\Delta_N d \to 0$ by
\be g_\omega(x,v_x) = \frac{\omega}{\Omega} + \frac{\omega \Delta_N}
{\Omega\left(\hbar\Omega-\Delta_N\right)} {\rm e}^{-\chi(x)}. 
\label{green_infinite_d}
\ee
The current in the presence of a slowly varying vector potential $A$, follows 
from Eq.\ (\ref{green_infinite_d}) after replacing $\omega$ by 
$\omega + iev_y A/\hbar c$ and inserting it into the current 
expression $j_y(x)= i e N_0 2 {\pi} T \sum_{\omega>0} 
\langle v_y g(x,v_x,v_y) \rangle$. In addition to the usual diamagnetic 
current $j_{\rm dia}  = -(c/4\pi \lambda^2)\, A$ 
[$\lambda = (mc^2/4\pi n e^2)^{-1/2}$ denotes the London length], we
obtain the paramagnetic current 
\be
j_{\rm para} \approx  \frac{c}{4\pi \lambda^2} {\rm e}^{-x/\alpha\xi_N^0}
    \alpha \frac{3 \Phi_0 }{2\pi \xi_N^0}\arctan \frac{e v_F A}{c \pi k_B T}, 
\label{jtpara} 
\ee
in the 
limit $ev_F A/c, k_BT \ll \Delta_N$. Here, $\Phi_0=\pi\hbar c/e$ denotes the 
flux quantum, $\xi_N^0 = \hbar v_F/2\Delta_N$ gives the extent of the 
$\pi$-states, and under the assumption $\Delta_N>\hbar v_F/d$ we have 
$\alpha=1$. At temperatures $k_B T \gg ev_F A/c$ the paramagnetic current 
$j_{\rm para}\sim (c/\lambda^2) (\Delta_N/k_BT)  A$ is 
linear in $A$ and $\propto 1/T$, a signature of the thermally smeared zero 
energy DOS peak, and competes with the diamagnetic current on the scale 
$\xi_N^0$. 
At $T\to 0$, Eq.\ (\ref{jtpara}) is nonlinear in the field and generates the 
spontaneous paramagnetic current. This paramagnetic interface current results 
from the energy splitting of the $\pi$-states in the field, 
$E\approx \pm e v_F A/c$, allowing the system to gain energy by shifting the 
DOS below the Fermi surface.
For $\Delta_N < \hbar v_F/d$, the paramagnetic current is reduced 
by the factor $\alpha=(\Delta_N d/\hbar v_F)<1$ in Eq.\ (\ref{jtpara}).
The surface current $I=\int j dx \sim \alpha^2 c\Phi_0/\lambda^2$ is in 
agreement with the current estimate $I_{\pi} \sim N_{\rm surf} e v_F$ based on 
the number of $\pi$-states at zero energy. Eq.\ (\ref{jtpara}) thus 
always produces a net paramagnetic response at low temperature and 
fields.\\
\vspace{-1.3cm}
\begin{figure}[htbp]
\psfig{figure=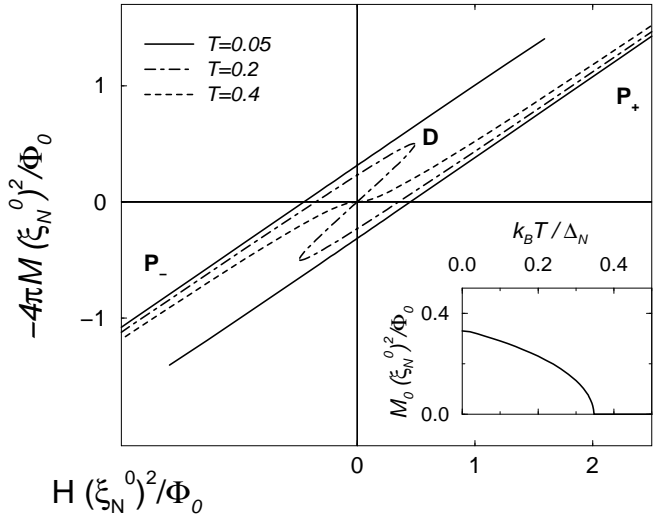,width=9.5cm,angle=-90}
\narrowtext
\vspace{0.5cm}
\caption{Magnetization $M(T,H)$ curve at various temperatures (in units of 
$\Delta_N / k_B$), for $\lambda=0.3d$ and $\xi_N^0=d$.
The two meta-stable branches $P_{\pm}$ exhibit a spontaneous magnetization  
in zero field, the diamagnetic branch $D$ is unstable. 
Inset: Zero field magnetization $M_0(T)$.}
\label{magfig}
\end{figure} 

The evaluation of the induced magnetization requires the self-consistent 
solution of Maxwell's equation $-\partial_x^2 A(x)= 4\pi j(x)/c$ together 
with the current functional $j[A(x)]$. The solution of the screening problem 
requires the full dispersive relation between $j(q)$ and $A(q)$, which in the 
proximity effect typically features a non-local current-field dependence on 
the 
scale $\xi_N^0$\cite{bbf}. Eq.\ (\ref{jtpara}) represents the long wavelength 
limit $q\to 0$. For simplicity, here we use Eq.\ (\ref{jtpara}) under the 
assumption of local response (accounting for the full non-locality does not
affect the qualitative nature of the final results, as we have checked 
numerically).

The magnetization curves $M(T,H)$, which follow from Eq.\ 
(\ref{jtpara}) are shown in Fig.\ \ref{magfig}. Approaching from large
fields, Fig.\ \ref{magfig} shows two paramagnetic branches $P_{\pm}$
with a linear diamagnetic slope exhibiting a spontaneous magnetization
in zero field. They result from the superposition of the paramagnetic
magnetization $M_0(T)$ and the Meissner response to the applied field
$H$.  As the field is decreased (increased) past $H=0$, the branch
$P_+$ ($P_-$) becomes meta-stable. The spontaneous magnetization
$M_0(T)$ appears below a second order transition point $T_c^M$ and
saturates at low temperatures, as shown in the inset of Fig.\ 
\ref{magfig}.  The magnetization curve includes a diamagnetic branch
$D$, which arises from the competition between the paramagnetic
instability and the thermal smearing and is thermodynamically
unstable.

In the following we give a semi-quantitative analysis of the
magnetization $M = \int dx \,M(x)/d$, first at zero temperature and
field [$M_0$], proceeding to finite temperatures [$M_0(T)$], and
finally including an applied magnetic field $H$ [$M(T,H)$]. The
boundary conditions are given by $A(0)=0$ and $\partial_x A(d)=H$. We
concentrate on the most relevant limit where $\xi_N^0, d \gg \lambda$.
At $T=0$, according to Eq.\ (\ref{jtpara}), the paramagnetic interface
current $j\sim \alpha c\Phi_0/\lambda^2\xi_N^0$ remains unscreened
until being matched by $j_{\rm dia}\sim -cA/\lambda^2$, producing a
vector potential $A\sim \alpha \Phi_0/\xi_N^0$ on the scale $\lambda$.
The vector potential $A$ saturates beyond $\lambda$, as the para- and
diamagnetic currents cancel each other. Assuming that the $\pi$-states
extend up to the outer metal surface ($\alpha<1$), the induced
magnetization $M = A(d)/4\pi d$ is given by 
\be M_0 \sim \alpha
\frac{\Phi_0}{\xi_N^0 d} \sim \frac{\Phi_0}{(\xi_N^0)^2}.
\label{m0}
\ee
We note that although the spontaneous currents increase as
$\Delta_N > \hbar v_F/d$ ($\alpha=1$) they are screened exponentially 
beyond the extent of the $\pi$-states in this limit, giving a magnetization 
$M_0 \sim (\Phi_0/\xi_N^0d) \exp[-(d-\xi_N^0)/\lambda]$. 
We assume $\alpha < 1$ in the following.

At finite temperature, the spontaneous magnetization is suppressed by
the factor $\arctan (e v_F A/c\pi k_B T)$, which itself depends
on the magnetization via $A \sim M d$, implying the implicit equation
\be \frac{M_0(T)}{M_0} \sim \arctan \frac{M_0(T) \alpha \Delta_N}
{M_0 k_B T}.
\label{condM}
\ee 
The spontaneous magnetization appears below a second order
transition at $k_B T_c^M \sim \alpha \Delta_N$,
saturating at low temperatures, as shown in the inset of Fig.\ 
\ref{magfig}. The transition temperature is equal in magnitude to the
energy splitting of the DOS peak $E \sim e v_F A/c \sim \alpha
\Delta_N$.
 
Under an applied magnetic field $H$, the Meissner current $j_{dia}$  
screens both the spontaneous interface current and the applied field.
At zero temperature we deal with a linear problem and the magnetization 
is given by the superposition $M(H) = M_0 + \chi H$ 
of the spontaneous magnetic moment $M_0$ and the Meissner response 
$\chi H$. As the temperature increases, $M_0(T)$ decreases and 
the meta-stable regime shrinks. At $T>T_c^M$ the spontaneous magnetization 
in zero field has disappeared, the signature of the paramagnetic currents 
remains, however, reducing the diamagnetic susceptibility $\chi$ at small
fields. At large temperature $T\!\gg\! T_c^M$ we recover the pure Meissner 
response.

Note that the two meta-stable branches $P_+$ and $P_-$ in the
magnetization curve, see Fig.\ \ref{magfig}, imply a first order
transition with changing field at $H=0$. The first order transition is
similar to the magnetic breakdown occurring in the same system at large
fields between the fully diamagnetic phase and a field penetration
phase\cite{alf}.  The rotation of the magnetic moments to the
energetically more favorable polarization will show the hysteretic
behavior typical for a first order transition. The transition from
$P_+$ to $P_-$ implies a paramagnetic slope in the thermodynamic
$dc$-magnetization $\langle M(T,H) \rangle$, which will link the
meta-stable solutions $P_{\pm}$ in Fig.\ \ref{magfig} and cross the origin at
$M(H=0)=0$. In summary, we find that on approaching $T_c^M$ from above, the 
diamagnetic susceptibility $\chi_{dc}=\langle M(T,H)\rangle/H$ is reduced, 
exhibiting a low-temperature reentrance. Below $T_c^M$, the 
spontaneous interface currents produce a net paramagnetic susceptibility 
$\chi_{dc}$.

In the following we discuss our results in the context of the
experiments by Mota and co-workers, who have measured the magnetic
response of normal-metal coated superconducting cylinders at low
temperatures\cite{motaprl,motapd}.  Recent studies have established a
quantitative understanding of the magnetic response of these samples
at higher temperatures, explaining the sensitivity of the screening to
small impurity concentration\cite{bbf,mmb} due to the nonlocality and
the magnetic breakdown at finite field\cite{alf}. The Nb-Ag and Nb-Cu
cylinders show an anomalous paramagnetic signal in the magnetic
response in the low-temperature -- low-field corner of the $H-T$ phase
diagram\cite{motaprl,motab}. A direct comparison with our theory
requires the magnetization curve $\langle M(T,H) \rangle$ which has
not yet been measured.  The observed $dc$-susceptibility $\chi_{dc}(T)$
as a function of temperature shows an increase at low
temperature\cite{frassanito}.  The measured $ac$-susceptibilities
$\chi_{ac}(T)$ and $\chi_{ac}(H)$ exhibit a reentrance both as
a function of temperature and field\cite{motaprl}.  The reentrance is
accompanied by an out-of-phase response signaling dissipation and by
hysteresis in the field dependence. These features are in qualitative
agreement with our results for the magnetization curve.  We find that 
theory and experiment agree in order of magnitude for $\alpha^2 \sim
0.1$, implying a transition temperature $T_c^M \sim 100 {\rm mK}$ and a
spontaneous magnetization $M_0 \sim 1 {\rm G}$.
A more quantitative comparison with experiment requires a
self-consistent treatment of the spontaneous currents with the pair
potential, accounting for the nonlocality of the current-field
relation and its sensitivity to disorder.

In conclusion, we have demonstrated that the inclusion of a finite
electron-electron repulsion in a proximity coupled normal-metal layer
naturally produces spontaneous interface currents leading to a
paramagnetic reentrance in the magnetic response: The sign change in
the coupling across the NS interface leads to the trapping of
$\pi$-states at the Fermi energy.  The frustrated NS junction relaxes
through the generation of spontaneous interface currents, inducing a
paramagnetic moment. A non-trivial issue remains the requirement that
the electron-electron interaction be repulsive at the low energy
scales involved.  Interesting consequences of this assumption have
been discussed in the context of the proximity effect \cite{deGennes}
and most recently in relation to the low temperature transport in
mesoscopic NS structures \cite{nazarov,petrashov}.  In fact, the noble
metal coatings used in the experiments of Mota and co-workers
\cite{motaprl} appear to be the most plausible candidates for a
repulsive electron-electron interaction. Turning the argument around,
in the light of our findings the experimental observation of a
paramagnetic reentrance can be taken as an indication of the presence
of a repulsive interaction.

We acknowledge fruitful discussions with C.\ Bruder, M.\ Dodgson, V.\ 
Geshkenbein, C.\ Honerkamp, A.\ Mota, B.\ M\"uller, 
M.\ Rice, G.\ Sch\"on, and M.\ Sigrist.

\end{multicols}

\end{document}